# Double O-Ne-Mg white dwarfs merging as the source of the Powerfull Gravitational Waves for LIGO/VIRGO type interferometers.

V.M.Lipunov [1,2]

[1] *M.V.Lomonosov Moscow State University, Physics Department, Leninskie gory, 1, Moscow, 119991, Russia*

[2] *M.V.Lomonosov Moscow State University, SAI, Universitetsky pr., 13, Moscow, 119234, Russia*

______________________________________________________________________

**Abstract:**
New strong non-spiralling-in gravitational wave (GW) source for LIGO/VIRGO detectors are proposed. Double O-Ne-Mg white dwarf mergers can produce strong gravitational waves with frequencies in the several hundreds Hz range. Such events can be followed by a Super Nova type Ia.

The discovery of gravitational waves (Abbot et al., 2016a,b,c) produced by a black-hole merger and predicted by the first Monte-Carlo code of the ScenarioMashine (Lipunov, Postnov, Prokhorov 1987a,b,c) confirmed the most general ideas about the evolution of baryonic matter in the Universe in the form of the evolution of binary stars (van den Heuvel and Heise, 1972; van den Heuvel and Loore, 1973; Tutukov and Yungelson, 1973, Yungelson and Tutukov, 1992; Lipunov et al., 2017). The next task is to discover mergers involving two neutron stars or a neutron star and a black hole, see references Clark et al. (1979), Lipunov (2016). However, we will show that there is yet another channel for the generation of powerful gravitational waves during the formation of a massive neutron star or a light black hole and this event should be accompanied by a SNIa explosion.

Because of the idea suggested by Iben and Tutukov (1984) and Webbink (1984) we now know that a merger of white dwarfs may result in an **SNIa** event. Moreover, **SNIa** observations in elliptical galaxies (Totani et al. (2008) demonstrate that white-dwarf mergers are the main channel for the formation of **SNIa** in elliptical galaxies, at least during after the first billion years of their life time (Lipunov, Panchenko, Pruzhinskaya, 2012).

The condition for explosion is that the total mass of white dwarfs must exceed the Chandrasekhar limit, see Masevich & Tutukov (1981); van den Heuvel (2011); Livio (2000):

$$M_{Ch} \sim 5.83\, \mu_e^{-2} \sim 1.4\, M_\odot,$$

where $\mu_e \approx 2$ is the average nucleon to electron ratio in the white dwarf.

White-dwarf merger must be accompanied by a gravitational-wave pulse (Cutler, Thorne, 2002). It was, however, pointed out that the low amplitude and low frequency (~1 Hz) of such a pulse compared to those of pulses accompanying NS and BH mergers these events are of little interest for LIGO type interferometers, at least in the years to come.

The point is that the overwhelming majority of merging white dwarfs whose merger results in SNIa explosion are the so called CO-type white dwarfs. After the merger an unstable CO-type dwarf forms, which, after the beginning of the collapse, ignites, breaks apart, and scatters as a result of a thermonuclear explosion leaving no remnant , see Iben,Tutukov (1984), van den Heuvel (2011). The maximum amplitude of the GW at the merger point can actually be estimated as **$h_0 \sim (Rg/R)^{5/2} \sim 10^{-5}$**, which is about 1000 times less than expected in the case of a double neutron star merger, resulting in a factor of $10^9$ smaller detection volume. Furthermore, the characteristic rotation frequency of a white dwarf should be, as we already wrote, well below the sensitivity curve of interferometers.

However, a merger of the more massive O-Ne-Mg dwarfs results in quite a different picture. After the merger of the dwarfs an O-Ne-Mg core may form with a mass exceeding 2.5 solar masses. Such a white dwarf will have huge angular momentum and the collapse may result in the development of a Dedekind type instability and formation of large quadrupole moment. Unlike the collapse of merged CO dwarfs that of merged O-Ne-Mg dwarfs does not result in complete disintegration (Nomoto 1984; Tominaga et al., 2008), and given such a large mass it would be natural for a massive neutron star or, more likely, a black hole to form. In this case the initial GW amplitude should be about $h_0 \sim 0.1-0.3$, which is comparable to the case of a usual merger of two relativistic stars.

The collapse of O-Ne-Mg dwarfs may also proceed via a different way, in accordance with the so called SD mechanism of supernova formation. An O-Ne-Mg dwarf may build up mass as a result of accretion in a binary system (usually with a red dwarf companion).

This is the so called Accretion Induced Collapse. However, this scenario results in a much smaller angular momentum of the final neutron star or black hole. An extreme white dwarf with a higher than critical mass is a very centrally concentrated object with a density contrast of ~1/50. Hence most of the angular momentum should be distributed in the exterior layers of the white dwarf whereas its more massive central part would have much smaller angular momentum and hence possesses a small quadrupole moment.

Thus a merger of two O-Ne-Mg dwarfs provides a competing channel for the generation of strong bursts of gravitational waves compared to NS and BH mergers. The evolutionary sequences:

$WD_{O-Ne-Mg} + WD_{O-Ne-Mg} \rightarrow NS + SNIa + GWB$
or
$WD_{O-Ne-Mg} + WD_{O-Ne-Mg} \rightarrow BH + SNIa + GWB$

depend on the Openheimer-Volkoff limit value. The collapse of a heavy no-solid-body-rotating O-Ne-Mg WD is a very complicated .

The maximal frequency is equal to 2 times the maximal spin frequency for a NS. The maximal spin frequancy is well known from millisecond pulsar observations and is equal about 700 Hz. This maximal frequency may be connected with the neutron stars equation of state (Lipunov & Postnov, 1984; Friedman et al., 1985). In the case of NS formation, the fast rotating pulsar can be observed.

Let us estimate the probability of these events. The mass ratio distributionof the main-sequence progenitors is **φ(q) = const** (Lipunov, Postnov, Prokhorov 1996, Fedorova et al. 2004), where **$0 < q=M_2/M_1 < 1$**.

In our case both components should have masses in the **$M_0 \sim 8\text{-}10\ M_\odot$** interval. The probability of the formation of two stars of similar mass is about ~10 times lower than the probability of the formation of stars that could in principle produce white dwarfs. However, we need two sufficiently massive stars. Such stars form in accordance with the Salpeter function:

$$dN/dM \sim M^{-\alpha}$$

Where **α** = 2.35 . Assuming all stars more massive than 3 solar masses to form CO white dwarfs sufficiently massive to produce a Type Ia supernova.

The relative probability of the formation of two O-Ne-Mg dwarfs compared to that of the formation of binary white dwarfs with total mass more than the Chandrasekhar limit is of about **$10^{-1}$**. We can therefore expect **$10^{-2}$** of all type Ia supernovae possibly to be strong gravitational wave sources.

The SNIa frequency at the current epoch, i.e. after about 10 Gyr (see Jørgensen et al., 1997; Totani. 2008; Lipunov et al., 2011), is ~ $1/300$ yr$^{-1}$ per $10^{11}$ solar masses. At the same time, the frequency of double NS mergers is ~ 1/10 000 - 1/30000 per $10^{11}$ solar mass (Lipunov et al., 1987; Grishchuk et al., 2001, Kalogera et al. 2004). Hence the collapse rate of binary O-Ne-Mg dwarfs may occur with about the same frequency as double neutron star mergers, and given that we are dealing with collapse into a black hole whose GW amplitude is twice higher than for neutron stars the corresponding detection rate may be even higher than the detection rate of double neutron star mergers.

I am grateful Alexander Tutukov, Edvard van den Huevel and Sergey Blinnikov for usefull remarks.